\documentclass[aps,prl,twocolumn,superscriptaddress,floatfix]{revtex4-1}

\usepackage{graphicx}
\usepackage{amsmath,amssymb}
\usepackage{hyperref}
\usepackage{datetime}
\usepackage[utf8]{inputenc}
\usepackage{braket}

\begin{document}

\title{Signatures of bath-induced quantum avalanches in a many-body--localized system}

\affiliation{Department of Physics, Harvard University, Cambridge, Massachusetts 02138, USA}
\affiliation{Institute for Theoretical Physics, ETH Zurich, 8093 Zurich, Switzerland}
\affiliation{Center for Computational Quantum Physics, Flatiron Institute, New York, New York 10010, USA}
\affiliation{Department of Physics, New York University, New York, New York 10003, USA}

\author{Julian~L\'eonard$^{1,*,\dagger}$}
\author{Sooshin~Kim$^{1,\dagger}$}
\author{Matthew~Rispoli$^1$}
\author{Alexander~Lukin$^1$}
\author{Robert~Schittko$^1$}
\author{Joyce~Kwan$^1$}
\author{Eugene~Demler$^{2}$}
\author{Dries~Sels$^{3,4}$}
\author{Markus~Greiner$^{1,\ddagger}$}

\begin{abstract} Strongly correlated systems can exhibit surprising phenomena when brought in a state far from equilibrium. A spectacular example are quantum avalanches, that have been predicted to run through a many-body--localized system and delocalize it. Quantum avalanches occur when the system is locally coupled to a small thermal inclusion that acts as a bath. Here we realize an interface between a many-body--localized system and a thermal inclusion of variable size, and study its dynamics. We find evidence for accelerated transport into the localized region, signature of a quantum avalanche. By measuring the site-resolved entropy we monitor how the avalanche travels through the localized system and thermalizes it site by site. Furthermore, we isolate the bath-induced dynamics by evaluating multipoint correlations between the bath and the system. Our results have fundamental implications on the robustness of many-body--localized systems and their critical behavior.\end{abstract}

\maketitle

\date{\today}

\twocolumngrid

One of the founding principles of statistical physics is that a generic macroscopic system can equilibrate on its own. This means that local fluctuations of energy, magnetization, or particle density can relax towards thermal equilibrium because interactions allow different parts of the system to serve as reservoirs to each other. This universal picture has been challenged by the idea of many-body localization (MBL), which suggests that systems with strong disorder can evade thermalization even in the presence of  interactions \cite{Alet2018,Abanin2019,Schreiber2015,Smith2016,Choi2016,RubioAbadal2019,Lukin2019,Lueschen2017a,Rispoli2019}. 

In one-dimensional systems, a stable MBL phase can be argued as follows: Matrix elements of local operators decay exponentially with separation between two points, whereas the density of states increases exponentially with the system size. For strong disorder, matrix elements can thus be argued to decay faster than the density of states increases, ultimately inhibiting relaxation. However, the existence of MBL remains a subject of debate \cite{Abanin2019a,Panda2020,Sierant2020,Suntajs2020,Suntajs2020a,Luitz2020,Kiefer2020,Kiefer2020a}, since it is unclear whether those conditions can actually be fulfilled. For instance, by introducing a small region with weak disorder, part of the system may be delocalized and thus give rise to local operators with non-exponential decay \cite{Agarwal2015,BarLev2015,Znidaric2016,Gopalakrishnan2016,Agarwal2017,Potter2015, Vosk2015,Gopalakrishnan2015,Weiner2019,Khemani2017,Khemani2017a,Weiner2019}. Those weakly disordered regions occur naturally in randomly disordered systems, when potential offsets on consecutive lattice sites accidentally coincide \cite{Griffiths1969,McCoy1969}.
The dynamics in MBL systems in the presence of a locally thermalizing region have been predicted to occur in so-called quantum avalanches, which imply those small islands grow by absorbing nearby disordered regions \cite{Nandkishore2017,Roeck2017,Luitz2017,Thiery2018,Crowley2020}. Under which conditions quantum avalanches can arise, run out of steam, or propagate without halt determines the ultimate fate of MBL at very long times. Their understanding is thus closely connected to discerning thermalization in interacting many-body systems.

\begin{figure}[h!b]
	\includegraphics[width=\columnwidth]{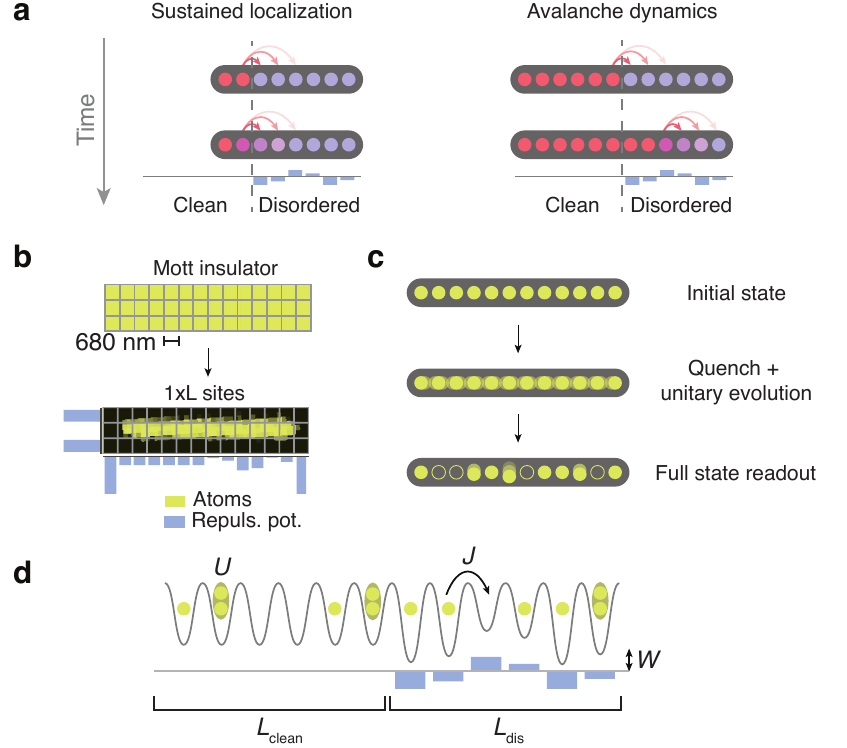}
	\caption{\textbf{Bath-induced quantum avalanches. a,} Two scenarios at an interface of a thermal bath (clean) and a localized (disordered) region: a weak bath penetrates logarithmically slow and localization remains robust (left), or an avalanche from a strong bath thermalizes the disordered region site by site (right). \textbf{b,} Fluorescence pictures of a two-dimensional Mott insulator at unity filling, and of the initialized one-dimensional system of $L$ sites. Projected optical potentials isolate the system and apply site-resolved offsets onto the disordered region (blue). \textbf{c,} The initial state is brought far from equilibrium through a quantum quench by abruptly enabling tunneling along all links, then evolved under the Hamiltonian, until we detect the site-resolved atom number with a fluorescence picture. \textbf{d,} The system's dynamics are governed by the Bose-Hubbard model with tunneling energy $J$ and on-site interaction energy $U$, extended by a disorder potential with amplitude $W$ in the disordered region.}
	\label{fig:cartoon}
\end{figure}

\begin{figure*}[t]
\centering
	\includegraphics[width=\textwidth]{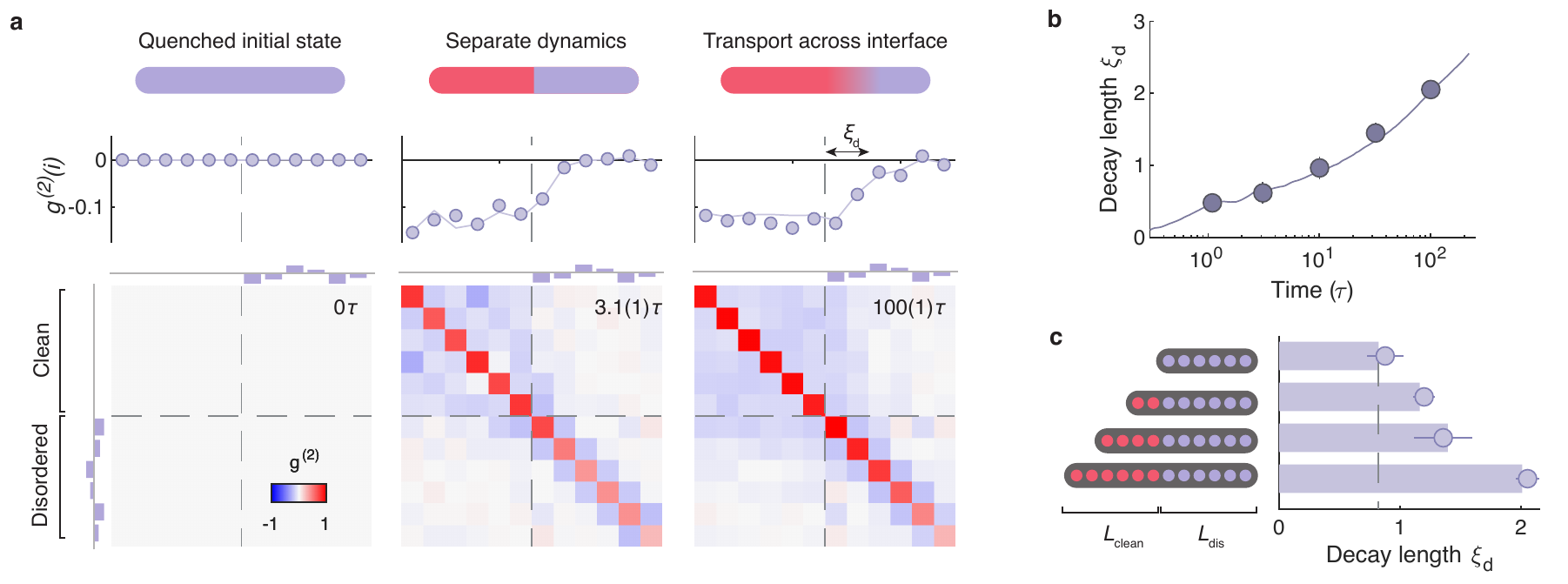}
	\caption{\textbf{Accelerated transport across the clean-disorder interface. a,} Density correlations for all pairs of sites in a system consisting of $L_\text{clean}=L_\text{dis}=6$ at disorder strength $W=9.1\,J$. After a quantum quench, an uncorrelated initial state (left) develops separate dynamics within each subsystem (center), followed by particle transport across the clean-disorder interface (grey dashed lines) for evolution times $\gg L_\text{clean},L_\text{dis}$ (right). Cuts show the total density correlations $g^{(2)}(i)$ of the clean region with site $i$ (i.e. average of top six rows, excluding diagonal entries), featuring homogeneous coupling among the clean sites, and exponentially decaying anti-correlations with the distance of the disordered site from the interface. \textbf{b,} The decay length $\xi_\text{d}$ of the total density correlations increases first logarithmically in time and accelerates at long evolution times. \textbf{c,} The decay length $\xi_\text{d}$ after an evolution time of $100\tau$ grows with $L_\text{clean}$, indicating improved particle transport into the disordered region. The data point at $L_\text{clean}=0$ and the dashed line show the localization length of an isolated MBL system. Solid lines (bars in panel c) show the prediction from exact numerics without free parameters. Error bars denote the s.e.m. (below the marker size in panel a).}
	\label{fig:g2}
\end{figure*}

Bath-induced relaxation dynamics can often be captured semi-classically in the context of Fermi's golden rule. In an isolated MBL system particle rearrangements are restricted to the length scales of the order of the localization length $\xi_\text{loc}$. The relaxation rate $\Gamma_i$ of a lattice site at distance $i$ coupled to the bath is captured by Fermi's golden rule $\Gamma_i = g_i^2 \rho_\text{bath}$. Here, the coupling for a relaxation process on site $i$ away from the bath leading to a transfer of energy or particles into the bath is set by $g_i \propto J e^{-i/\xi_\text{loc}}$. The density of states in the thermal region is exponential in its size, i.e. $\rho_\text{bath} \propto J^{-1} e^{\alpha L_\text{bath}}$ with a constant $\alpha$. This model implies that site $i$ shows relaxation after a time $T_i = 1/\Gamma_i$, or equivalently, after an evolution time $T$ we expect relaxation on the sites up to the distance $d_\text{FGR}(T) \sim \xi_\text{loc} \log (J^2 \rho_\text{bath} T)$. In conclusion, within a perturbative description MBL remains robust against a local bath, with a bath penetration into the MBL region that increases only logarithmically in time. Quantum avalanches, however, are predicted to emerge from dynamics beyond Fermi's golden rule. As the bath begins to delocalize neighboring disordered sites, the size of the thermalizing bath expands, leading to an increase in its density of states. 

In this work we explore the dynamics of an MBL system coupled to a thermal bath (Fig.\,\ref{fig:cartoon}). We observe phenomena that suggest the presence of non-perturbative avalanche processes, while other features of dynamics can be explained using the perturbative Fermi's golden rule. Our experimental protocol starts by preparing a Mott-insulating state with one \textsuperscript{87}Rb atom on each site of a two-dimensional optical lattice (Fig.\,\ref{fig:cartoon}b). The system is placed in the focus of a high-resolution imaging system through which we project site-resolved repulsive potentials on individual lattice sites. We isolate a one-dimensional system of $L$ lattice sites from the Mott insulator and add potential offsets to the lattice sites. At this point, the system remains in a product state of one atom per lattice site. We then perform a quantum quench by abruptly reducing the lattice depth (Fig.\,\ref{fig:cartoon}c). The subsequent non-equilibrium dynamics are described by the Bose-Hubbard Hamiltonian: 

\begin{equation*}
\begin{aligned}
	\hat{\mathcal{H}} &=  -J \sum_i \left(\hat{a}_i^\dagger\hat{a}_{i+1} + h.c.\right) \\
	&+ \frac{U}{2} \sum_i\hat{n}_i \left( \hat{n}_i - 1\right) + W\sum_{i\in L_\text{dis}} h_i\hat{n}_i\text{,}
\end{aligned}
\end{equation*}

\noindent where $\hat{a}^\dagger_i$ ($\hat{a}_i$) is the creation (annihilation) operator for a boson on site $i$, and $\hat{n}_i= \hat{a}^\dagger_i \hat{a}_i$ is the particle number operator. The first term describes the tunneling between all neighboring lattice sites, and the second term represents the on-site repulsive interactions. The last term introduces a site-resolved energy offset. We set $h_i=0$ for all lattice sites in the clean region of size $L_\text{clean}$, whereas the energy offsets in the disordered region of size $L_\text{dis}$ follow a quasi-periodic disorder distribution $h_i=\cos(2\pi\beta i+\phi)$ with $1/\beta \approx 1.618$, phase $\phi$ and amplitude $W$. The quasi-periodic distribution avoids nearby lattice sites to coincidentally have similar energy offsets, which inhibits the presence of secondary rare regions within the disordered region \cite{Setiawan2017}. After a variable evolution time, we read out the site-resolved atom number by fluorescence imaging. The applied unitary evolution preserves the initial purity of $99.1(2)\%$ per site \cite{Kaufman2016,Lukin2019}. All observables are disorder-averaged by realizing potential with different $\phi$. The tunneling time $\tau=\hbar/J =4.3(1)\,\text{ms}$ (with the reduced Planck constant $\hbar$), the interaction strength $U=2.87(3)\,J$, and the number of disordered sites $L_\text{dis}=6$ remain constant in all experiments. 

We first use the full site-resolved readout of our microscope to investigate the local transport dynamics in the system. The connected density-density correlations $\langle \hat{n}_i \hat{n}_j \rangle_c = \langle \hat{n}_i \hat{n}_j \rangle - \langle \hat{n}_i\rangle\langle \hat{n}_j\rangle$ detects correlations between the particle numbers on site $i$ and $j$ \cite{Rispoli2019}. Negative values of $\langle \hat{n}_i \hat{n}_j \rangle_c$ signal anti-correlated density fluctuations, and thus particles motion between the involved sites (Fig.\,\ref{fig:g2}a). In the following, we consider a system with $L_\text{clean}=6$ at disorder strength $W=9.1\,J$ after different evolution times $T$ after the quantum quench. At the beginning of the evolution ($T=0\tau$), we do not detect any correlations, because the initial state is a product state. After short evolution times ($T\lesssim\tau L$), we observe the buildup of spatially dependent anti-correlations in the system. Within the clean region all lattice sites develop mutual anti-correlations, signaling delocalizing particles. In contrast, the anti-correlations in the disordered region remain short-ranged, indicating localized particles. At this time, we do not detect significant anti-correlations between the clean and the disordered region. 

The situation changes for long evolution times ($T\gg \tau L$), where the correlations in the clean region have spread out evenly among all pairs of lattice sites, signaling homogeneously delocalized particles. Furthermore, we observe the buildup of anti-correlations between lattice sites in the clean and the disordered region, evidence for transport dynamics across the interface. Each of the disordered sites is equally anti-correlated to all clean sites, which suggests that the clean region acts as a homogeneous bath for the disordered region. Motivated by this picture, we extract the total correlations of the clean region $g^{(2)}(i)=\sum_{j\in L_\text{clean}} \langle \hat{n}_i \hat{n}_j\rangle_\text{c}$ by taking the sum of the correlations of each site with all clean sites (Fig.\,\ref{fig:g2}b cute). The results show a decay with distance from the clean region, in agreement with the Fermi golden rule picture of exponentially decaying couplings between bath and MBL. 

While a static bath spectrum causes bath correlations to penetrate MBL logarithmically in time, a signature of the quantum avalanche is an accelerated increase, faster than logarithmically in time. In order to test this picture, we quantify the correlation decay into the disordered region by measuring the average distance $\xi_\text{d}=\sum_{i\in L_\text{dis}} i g^{(2)}(i)$ from the clean region over which anti-correlations form (Fig.\,\ref{fig:g2}b). At short times the decay length $\xi_\text{d}$ increases logarithmically in time, but accelerates at long evolution times --- signature for the emergence of a quantum avalanche. 

\begin{figure}[t]
	\includegraphics[width=\columnwidth]{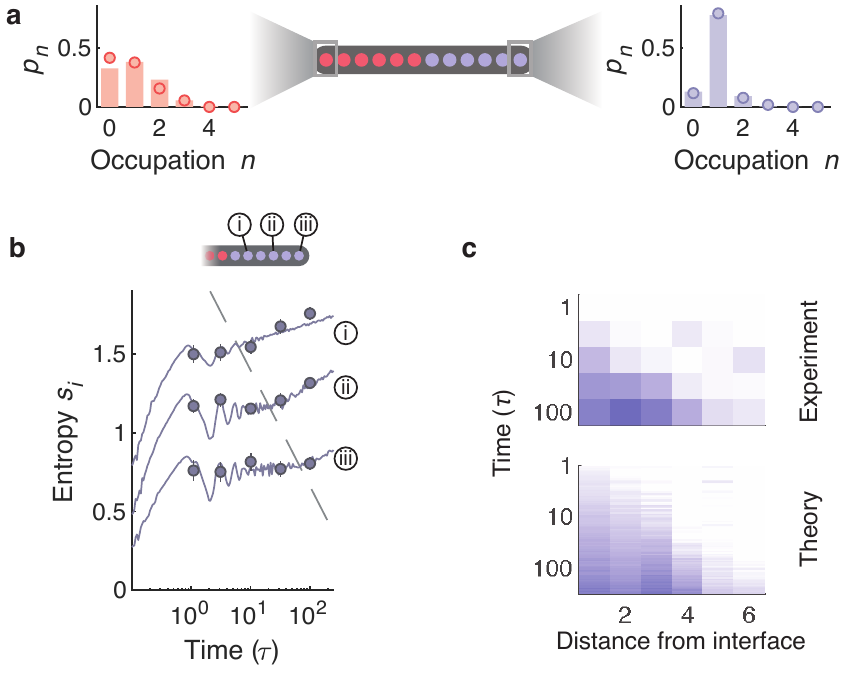}
	\caption{\textbf{Site-resolved thermalization dynamics. a,} The atom number probability distribution for the edge sites in the clean region (left) and the disordered region (right), measured after $100\tau$ in a system consisting of $L_\text{clean}=L_\text{dis}=6$ at disorder strength $W=9.1\,J$. \textbf{b,} Local entropy per particle $s_i=-\sum_n p_n\log p_n/\langle \hat{n}_i\rangle$ extracted from the atom number distribution on site $i$. The entropy grows after a stationary evolution whose length depends on the distance from the interface (indicated by the grey dashed line). Traces are vertically offset for better readability. \textbf{c,} Local entropy $s_i$ (offset by $s_i(T=1\tau)$) for all disordered sites. Solid lines (bars in panel a) show the prediction from exact numerics without free parameters. Error bars denote the s.e.m. (below the marker size in panel a).}
	\label{fig:svn}
\end{figure}

\begin{figure}[t]
	\includegraphics[width=\columnwidth]{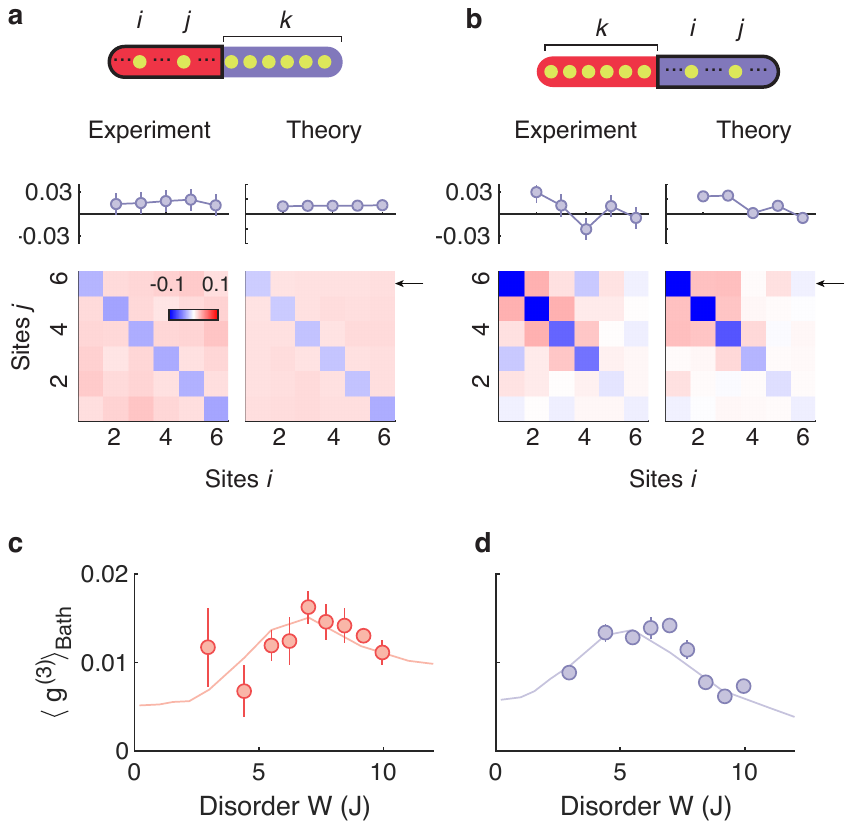}
	\caption{\textbf{Bath-induced many-body correlations. a,} Three-point correlations  $\langle \hat{n}_i \hat{n}_j \hat{n}_k\rangle_c$ among pairs of clean sites $i$, $j$ and one disordered site $k$ (summed over all disordered $k$) in a system with $L_\text{clean}=L_\text{dis}=6$ at disorder strength $W=9.1\,J$ and evolution time $T=100(1)$. Cuts across the site $j=6$ (arrows) show nonzero entries for all sites, evidence for multi-particle entanglement between all sites in the clean region with the disordered sites. The flat distribution visualizes the homogeneous coupling to the disordered region. \textbf{b,} Correlations $\langle \hat{n}_i \hat{n}_j \hat{n}_k\rangle_c$ among pairs of disordered sites $i$, $j$ and one clean site $k$ (summed over all clean $k$) vary strongly with the chosen lattice sites, and decrease with the distance from the clean region. The presence of multi-point correlations demonstrates non-perturbative dynamics: delocalization is driven through many-body processes between the disordered region and the clean region. \textbf{c,} We average over all off-diagonal sites and find a maximum for intermediate disorder for the MBL-bath entanglement. \textbf{d,} The total multi-point correlations among disordered sites with the bath show a similar maximum at slightly lower intermediate disorder. Solid lines show the prediction from exact numerics without free parameters. Error bars denote the s.e.m.}
	\label{fig:g3}
\end{figure}

The size $L_\text{clean}$ determines the number of degrees of freedom of the initial thermal region, and thus the spectral density of the thermal bath. While a bath of small number of degrees of freedom can only couple to disordered sites at distances on the order of the localization length $\xi_\text{loc}$, larger baths are expected to significantly exceed this length scale. The perturbative picture predicts that $\xi_\text{d}\propto \xi_\text{loc} \log(J\rho_\text{bath})\propto \xi_\text{loc} \times L_\text{clean}$, therefore a deviation from this proportionality can be regarded as evidence for non-perturbative dynamics in form of avalanches. In order to investigate this effect, we realize systems with different $L_\text{clean}$, while keeping $L_\text{dis}=6$ constant (Fig.\,\ref{fig:g2}c). For each system size, we characterize the particle transport by measuring $\xi_\text{d}$ after an evolution time of $100(1)\tau$. Our results show an increasing value of $\xi_\text{d}$ for larger $L_\text{clean}$. The enhanced $\xi_\text{d}$ for $L_\text{clean}=6$ suggests the presence of a quantum avalanche in the system. 

We next examine the local thermalization dynamics in a system with $L_\text{clean}=L_\text{dis}=6$. The site-resolved full atom number readout enables us to measure the atom number distribution on a local level (Fig.~\ref{fig:svn}a). Lattice sites in the clean region show a distribution corresponding to a thermal ensemble, whereas lattice sites in the disordered region show a distribution with enhanced probability for one particle, the initial state of the system. We quantify the site-resolved thermalization dynamics with the entropy per particle $s_i=-\sum_{n_i}{p(n_i)} \log p(n_i)/\langle \hat{n}_i\rangle$ on site $i$ from the atom number distributions. We observe reduced thermalization dynamics of the disordered sites with increasing distance from the interface (Fig.~\ref{fig:svn}b). Moreover, the data suggest that the dynamics are first stationary until thermalization sets in with a delay that is exponential in the site's distance from the interface. This picture is confirmed by our exact numerical calculations. 

The signatures for quantum avalanches imply that many-body processes drive the long-term dynamics of the system. We investigate this effect through multipoint correlations \cite{Kubo1962, Rispoli2019}. The presence of non-zero three-point connected correlations $\langle \hat{n}_i \hat{n}_j \hat{n}_k\rangle_c$ signals the presence of entanglement among all involved lattice sites, which cannot be explained by lower order processes. We start by evaluating the connected correlations $\langle \hat{n}_i \hat{n}_j \hat{n}_\text{dis}\rangle_c$ among two clean lattice sites $i$, $j$ and a disordered site $k$, summed over all possible $k$ (Fig.\,\ref{fig:g3}a). The correlations are non-zero across the clean region, and their homogeneous distribution indicates that all clean sites contribute equally to the delocalization in the disordered region. In contrast, when evaluating the connected correlations $\langle \hat{n}_i \hat{n}_j \hat{n}_\text{clean}\rangle_c$ among two disordered sites $i$, $j$ and a clean site $k$, averaged over all possible $k$ (Fig.\,\ref{fig:g3}b), the data show a strong dependence on the involved disordered sites. Close to the interface we find strong correlations, whereas they are absent for distant sites. We quantify the presence of many-body correlations at different disorder strengths and find a maximum at intermediate strengths (Fig.\,\ref{fig:g3}c,d), close to the estimated critical point of the system \cite{Rispoli2019}. 

In conclusion, we experimentally studied signatures of quantum avalanches in an MBL system, set in motion by a thermal inclusion. We observed an accelerated intrusion of the bath in the MBL system, its evolution to thermal equilibrium site after site, and the many-body entanglement between the two subsystems. By varying the size $L_\text{clean}$, we studied the emergence of quantum avalanches for increased number of degrees of freedom of the bath. In future, our experiments can be readily extended in many ways. For example, one could more systematically study the fate of quantum avalanches as a function of bath size and localization length. By increasing both the system size of the disordered region, one could explore the interplay at intermediate disorder strengths in a quantitive way through its scaling behaviour, i.e. by increasing the system size at constant ratio of $L_\text{clean}$ and $L_\text{dis}$, which may provide insight into the critical behaviour of the transition. An interesting extension would also be the influence of the statistical distribution of the disorder on the critical behaviour of the system.

We acknowledge fruitful discussions with K.~Agarwal, V.~Khemani, M.~Knap, M.~Lebrat and J.~Marino. We are supported by grants from the National Science Foundation, the Gordon and Betty Moore Foundations EPiQS Initiative, an Air Force Office of Scientific Research MURI program, an Army Research Office MURI program, the Swiss National Science Foundation (J.~L.), and the NSF Graduate Research Fellowship Program (S.~K.). 

$^{*}$ current address: Vienna Center for Quantum Science and Technology, Atominstitut, TU Wien, Vienna, Austria; $^\dagger$ These authors contributed equally to this work; $\ddagger$ email: greiner@physics.harvard.edu

\bibliography{bib}

\section{Supplementary information}

\subsection{Experimental sequence}

\textbf{Mott insulator preparation.} All described experiments start with a Bose-Einstein condensate of bosonic $^{87}$Rb atoms in the $|F=1, m_F=-1\rangle$ hyperfine state. This ultracold gas is loaded into a single 2D plane of a deep lattice along the vertical direction with lattice constant $1.5\mu m$ at laser wavelength $760\,\text{nm}$. This lattice stays on for the remainder of the experiment. We use an attractive dimple potential to isolate a controlled number of atoms from the 2D gas and load them into the center of a repulsive ring-shaped potential, created from a second laser beam at wavelength $760\,\text{nm}$. At this point the atoms form a two-dimensional superfluid with harmonic in-plane confinement. We then ramp up further laser beams at wavelength $760\,\text{nm}$ over $250\,\text{ms}$ to create a repulsive two-dimensional square lattice with lattice constant $a=680\,\text{nm}$ in both directions and lattice depth $45E_\text{r}$, where $E_\text{r}=h^2/(2m a^2)=h \times 1.1\,\text{kHz}$ is the recoil energy of a $^{87}$Rb atom of mass $m$. 

\textbf{Initial state preparation.} We use two digital micro-mirror devices (DMD) to project repulsive potentials onto the Mott insulator. The DMDs are placed in the Fourier plane with respect to the atoms, which allows us to project diffraction limited arbitrary potentials that correct for optical wavefront aberrations in the imaging system \cite{Zupancic2016}. We optically confine a single chain of $L=L_\text{clean}+L_\text{dis}$ lattice sites within the Mott insulator's unity-filling shell, and subsequently ramp down the power of the optical lattice. We use a repulsive deconfining beam to eject all atoms outside the confinement potential, while each atom within the projected confinement potential remains pinned on its lattice site. We then ramp the lattice back to $45E_\text{r}$ and remove the confining DMD potential. After post-selecting for the atom number $N=L$, this procedure results in an initial state of  $99.1(2)\,\%$ fidelity per site. 

\textbf{Quantum quench and state evolution.} We use the first DMD to project a ``wall-potential'' on the adjacent sites around the one-dimensional system, which provides a box-like confinement. This potential is registered to the position of the optical lattice and defines the size of the one-dimensional system. We simultaneously use the second DMD to project a custom, quasi-periodic disorder potential onto the disordered region of the system. The disorder strength $W$ is tuned by the intensity of the DMD potential. The quantum quench is initiated by lowering the lattice depth along the one-dimensional system from  $45E_\text{r}$ to $8E_\text{r}$. After a variable evolution time we freeze the dynamics by ramping the lattice back to $45E_\text{r}$. 

\textbf{Full quantum state read out.} We first let the atom populations located on individual lattice sites expand into independent tubes and use fluorescence imaging with an optical molasses beam to perform a site-resolved atom number measurement. The expansion step before the imaging procedure is employed to avoid parity projection during the imaging process. We subsequently post-select our data by excluding any images which do not contain the correct total number of atoms. The error in postselection, that is the fraction of falsely post-selected snapshots due to the finite readout fidelity, is $<0.1\,\%$ for all the experiments, small compared to the statistical error in the data. 

\subsection{Calibration of Hamiltonian parameters}

The calibration procedure for the Bose-Hubbard parameters is identical to the one described in \cite{Lukin2019}. We obtain $J=h\times 37.5(1)\,\text{Hz}$ and $U=h\times 107(1)\,\text{Hz}$. 

\subsection{Multi-point correlations}

Generically, a $n^\mathrm{th}$ order correlation function can be measured from a set of operators $\mathcal{O}_i$ by their joint expectation value $\langle \prod_{i=1}^n \mathcal{O}_i \rangle = \langle \mathcal{O}_1 \mathcal{O}_2 ... \mathcal{O}_n \rangle$. However, this joint expectation value captures two kinds of information: ``disconnected'' correlations that exist at $n^\mathrm{th}$ order due to existing lower order correlations, and ``connected'' correlations that only exist at order $n$ and can't be described by factorization into correlations of lower order \cite{Kubo1962}.
 
In the two-point case, this would mean comparing the measured value of $\langle \mathcal{O}_i \mathcal{O}_j \rangle$ to the product of their individual expectation values $\langle \mathcal{O}_i \rangle \langle \mathcal{O}_j \rangle$. The ``connected'' part of the correlation between $i$ and $j$ is  defined as the correlations that remain after removing the contributions from factorization into smaller groups. This motivates the definition of $\langle \mathcal{O}_i \mathcal{O}_j \rangle_\text{c} =\langle \mathcal{O}_i \mathcal{O}_j \rangle - \langle \mathcal{O}_i \rangle \langle \mathcal{O}_j \rangle$.
 
 For a three-point connected correlation function, we must subtract out contributions that come from connected two-point correlations that can look like three-point correlations when randomly combined with a residual 1-point correlation. This is how the connected three-point correlation function is defined in the main text for the on-site number operator $\hat{n}_i$. 
 \begin{flalign*}
\langle \mathcal{O}_i \mathcal{O}_j \mathcal{O}_k\rangle_\text{c} = &{} \langle \mathcal{O}_i \mathcal{O}_j \mathcal{O}_k \rangle \\
 - &{} G_\text{c}^{(2)}(i,j) \langle \mathcal{O}_k \rangle - G_\text{c}^{(2)}(i,k) \langle \mathcal{O}_j \rangle  - G_\text{c}^{(2)}(j,k) \langle \mathcal{O}_i \rangle \\
      - &{} \langle \mathcal{O}_j \rangle \langle \mathcal{O}_j \rangle \langle \mathcal{O}_k \rangle
\end{flalign*}

Higher order multi-point correlations can be constructed in a similar way \cite{Rispoli2019}. 
 
\subsection{Numerical calculations}
The experimental studied system sizes have Hilbert space dimensions of up to $1.3\times 10^6$ ($L=12$, N=12). Due to the non-equilibrium evolution and the disorder, matrix diagonalization for such systems is computationally challenging. Instead, we implement an exact numerical integration of Schr\"{o}dinger's equation $\ket{\psi(t)}=e^{-i\hat{H}t/\hbar}\ket{\psi_0}$ based on the Krylov-subspace method~\cite{EXPOKIT}. This method provides an memory- and CPU-run-time efficient way to numerically compute the time evolution while achieving high, controlled precision. All numerical calculations are averaged over 200 different realizations of the quasi-periodic potential. The computations are performed on the Harvard Odyssey computing cluster (for specifications see: https://www.rc.fas.harvard.edu/odyssey/).

\subsection{Data Analysis}
For all experiments we average over 197 patterns of quasi-periodic potentials, each with a different phase $\phi$ of the quasi-periodic potential. The data are taken from a running average over those patterns by randomly sampling a given number of realizations and treating them as independent measurements of the same system. 

We extract the decay length $\xi_d$ by computing the first moment of the non-local density-density correlations $\xi_d=\sum_i i\langle\hat{n}_i\hat{n}_j\rangle_c$.

The single-site entropy in Fig. 3a is extracted from the edge sites. The edge sites are most insensitive to the dynamics at the clean-disorder interface and therefore allow for a fair indicator for thermalization \cite{Khemani2017}. 

Error bars are computed by resampling the set of snapshots with replacement (bootstrapping).

The number of samples for each experiment is summarized in the following table:

\begin{table}[h]
{\renewcommand\arraystretch{1.25}
\begin{tabular}{rl|l|l|l|} %l|l|
Figure & \multicolumn{2}{p{6cm}}{Number of samples} \\ \hline
	2a,b & \multicolumn{2}{p{6cm}}{\raggedright 199 ($0\tau$), 86 ($1\tau$), 242 ($3.1\tau$), 294 ($10\tau$), 315 ($31.9\tau$), 456 ($100\tau$)} \\ 
	2c & \multicolumn{2}{p{6cm}}{\raggedright 456 ($L_\text{clean}=0$), 835 ($L_\text{clean}=2$), 134 ($L_\text{clean}=4$), 456 			($L_\text{clean}=6$)} \\ 
	3a & \multicolumn{2}{p{6cm}}{\raggedright 835 ($100\tau$)} \\ 
	3b,c & \multicolumn{2}{p{6cm}}{\raggedright same samples as for Fig. 2a,b} \\
	4a,b & \multicolumn{2}{p{6cm}}{\raggedright 456} \\ 
	4c,d & \multicolumn{2}{p{6cm}}{\raggedright 85 ($W=2.9\,J$), 71 ($W=4.4\,J$), 553 ($W=5.5\,J$), 179 ($W=6.2\,J$), 198 ($W=7.0\,J$), 191 ($W=7.7\,J$), 200 ($W=8.4\,J$), 623 ($W=9.1\,J$), 237 ($W=9.6\,J$)} \\ 
\end{tabular}}
\end{table}

\end{document}